\documentclass{PoS_arXiv}

\usepackage{amsmath,amssymb,amsfonts,graphicx,cite,colordvi,soul,stmaryrd,color}

\providecommand\Code[1]{\ensuremath{\texttt{#1}}}
\providecommand\Var[1]{\ensuremath{\mathit{#1}}}
\providecommand\Vi{\Var{i}}
\providecommand\Vj{\Var{j}}
\providecommand\Vt{\Var{t}}
\providecommand\Vg{\Var{g}}
\providecommand\Vgp{\Var{g'}}
\providecommand\Vh{\Var{h}}
\providecommand\Vhp{\Var{h'}}
\providecommand\Vs{\Var{s}}
\providecommand\Vsp{\Var{s'}}
\providecommand\Vo{\Var{o}}
\providecommand\Vn{\Var{n}}
\providecommand\Vnp{\Var{n'}}
\providecommand\Vc{\Var{c}}
\providecommand\Vcp{\Var{c'}}
\providecommand\Vu{\Var{u}}

\providecommand\opt[2]{\rlap{\raise .7ex\hbox{\small #1}}%
  \lower .9ex\hbox{\vphantom{g}\small #2}}
\providecommand\bbarb{\textrm{[}bar\textrm{]}}
\providecommand\rI{$|$}

\providecommand\ri{\mathrm{i}}
\providecommand\re{\mathrm{e}}
\renewcommand\Re{\mathop{\mathrm{Re}}}
\renewcommand\Im{\mathop{\mathrm{Im}}}

\providecommand\ReDiag{\mathop{%
  \raise .5pt\hbox{[}%
  \widetilde{\mathrm{Re}}%
  \raise .5pt\hbox{]}}}
\providecommand\ReOffDiag{\mathop{%
  \raise .5pt\hbox{$\llbracket$}%
  \widetilde{\mathrm{Re}}%
  \raise .5pt\hbox{$\rrbracket$}}}

\providecommand\DRbar{\ensuremath{\smash{\overline{\mathrm{DR}}}}}

\providecommand\matr[1]{\mathbf{#1}}
\providecommand\mati[1]{\bigl(#1\bigr)}

\providecommand\SW{s_\mathrm{w}}
\providecommand\CW{c_\mathrm{w}}
\providecommand\MW{M_W}
\providecommand\MZ{M_Z}
\providecommand\MA{M_A}
\providecommand\MHp{M_{H^\pm}}
\providecommand\mf[1]{m_{f_{#1}}}

\providecommand\Ab{A_b}

\providecommand\Sf{{\tilde f}}

\providecommand\Stop{{\tilde t}}

\providecommand\gl{{\tilde g}}
\providecommand\mgl{m_\gl}
\providecommand\phigl{\varphi_\gl}

\providecommand\cH[1]{\mathcal{H}_{#1}}

\providecommand\ino[1]{\tilde\chi_{#1}}

\providecommand\chapm[1]{\ino{#1}^\pm}

\providecommand\mcha[1]{m_{\chapm{#1}}}
\providecommand\neu[1]{\ino{#1}^0}

\providecommand\refta[1]{Table~\ref{#1}}

\providecommand\citere[1]{Ref.~\cite{#1}}
\providecommand\citeres[1]{Refs.~\cite{#1}}

\providecommand\lbrac{\symbol{123}}
\providecommand\rbrac{\symbol{125}}
\providecommand\Brac[1]{\lbrac#1\rbrac}

\renewcommand\arraystretch{1.2}
\renewcommand\tabcolsep{8pt}
\renewcommand\arraycolsep{8pt}

\title{Fully Automated Calculations in the complex MSSM}

\ShortTitle{Automated cMSSM Calculations}

\author{Thomas Hahn\\
        Max-Planck-Institut f\"ur Physik, %(Werner-Heisenberg-Institut),
        F\"ohringer Ring 6, D--80805 M\"unchen, Germany\\
        E-mail: \email{hahn@feynarts.de}}

\author{\speaker{Sven Heinemeyer}\\ %\thanks{A footnote may follow.}\\
      Instituto de F\'isica de Cantabria (CSIC-UC), E-39005 Santander, Spain\\
      E-mail: \email{Sven.Heinemeyer@cern.ch}}

\author{Federico von der Pahlen\\
        Instituto de F\'isica, Universidad de Antioquia, Calle 70 No.~52-21,
        Medell\'in, Colombia\\
        E-mail: \email{fp@gfif.udea.edu.co}}

\author{Heidi Rzehak\\
        Albert-Ludwigs-Universit\"at Freiburg, 
        Physikalisches Institut, D--79104 Freiburg, Germany\\
        E-mail: \email{Heidi.Rzehak@cern.ch}}

\author{Christian Schappacher\\
        Institut f\"ur Theoretische Physik,
        Karlsruhe Institute of Technology, 
        D--76128 Karlsruhe, Germany (former address)\\
        E-mail: \email{schappacher@kabelbw.de}}

\abstract{We review recent progress towards automated higher-order
calculations in the MSSM with complex parameters (cMSSM).
The consistent renormalization of all relevant sectors of the cMSSM
and the inclusion into the {\tt FeynArts}/{\tt FormCalc} framework has
recently been completed. Some example calculations applying this
framework are briefly discussed. These include two-loop corrections to
cMSSM Higgs boson masses as well as partial decay widths of electroweak
supersymmetric particles decaying into a Higgs boson and another
supersymmetric particle. 
}

\FullConference{ Loops and Legs in Quantum Field Theory - LL 2014,\\
		 27 April - 2 May 2014 \\
		 Weimar, Germany }

\include{paperdef}
\graphicspath{{figs/}}

\newcommand{\MHexp}{125.6 \gev}

\begin{document}

\section{Introduction}

Two of the most important goals of the experiments at the Large Hadron
Collider (LHC) are to identify the origin of electroweak symmetry
breaking (EWSB), and to search for physics effects beyond the Standard Model
(SM). The spectacular 
discovery of a Higgs-like particle 
with a mass around $\sim \MHexp$, which was announced by ATLAS
and CMS~\cite{ATLASdiscovery,CMSdiscovery}, marks a milestone of an
effort that has been ongoing for almost half a century and opens a new
era of particle physics.  
Within the experimental uncertainties the properties of the newly
discovered particle are in agreement with the predictions of the SM Higgs
boson~\cite{LHCP-ATLAS,LHCP-CMS}. However, the uncertainties still leave
room for contributions from non-SM degrees of freedom, see, e.g.,
\citeres{HiggsCouplings,LHCP-maggie,Hcoup-rev} for a recent combination and
reviews. 
The prime task now is to study the properties of the discovered
new particle in detail and to investigate whether there are
significant deviations from the SM predictions, which would point
towards physics beyond the SM.

The extent to which the results of the Higgs searches at the LHC can 
discriminate between the SM and possible alternatives depends both on
the experimental precision with which the properties of a possible
signal can be determined and on the detailed nature of the mechanism for
EWSB that is actually realized in nature.
One of the leading candidates for
physics beyond the SM (BSM) is supersymmetry (SUSY), which doubles the
particle degrees of freedom by predicting two scalar partners for all SM
fermions, as well as fermionic partners to all bosons. The most widely
studied SUSY framework is the Minimal Supersymmetric Standard Model
(MSSM)~\cite{mssm}, which keeps the number of new
fields and couplings to a minimum. The MSSM Higgs sector 
contains two Higgs doublets, which at the tree-level leads to
a physical spectrum consisting of two 
$\cp$-even, $h, H$, one $\cp$-odd, $A$, and two charged Higgs bosons,
$H^\pm$.

In order to investigate the impact of the Higgs search results at
the LHC on possible scenarios of new physics, precise theoretical
predictions both within the SM and possible alternatives of it are
needed. In particular, if small deviations from the SM predictions are
probed it is crucial to treat the considered model of new physics at the
same level of precision to enable an accurate analysis and comparison. 
In the MSSM Higgs sector higher-order contributions are known to give 
numerically large effects (see, e.g.,
\citeres{MSSMHiggsRev,MSSMHiggsRev2}).
For many observables it is therefore 
necessary to include corrections beyond leading order in the
perturbative expansion to obtain reliable results. 
The calculation of loop diagrams,
often involving a large number of fields, is a tedious and
error-prone task if done by hand. This is true in particular for
BSM theories where the number of fields is significantly
increased. For one-loop calculations, as will be the focus in the
following, computer methods with a high degree of automatization have
been devised to simplify the work. However, most of the available tools
so far have focused on calculations either in the SM or the MSSM with 
external SM particles.

Here we review renormalization of the MSSM including complex parameters
(cMSSM) and the corresponding implementation as a model
file~\cite{mssmct} into the  
{\tt FeynArts}~\cite{FeynArts,FA-MSSM}/{\tt FormCalc}~\cite{FormCalc}
framework. This implementation allows for automated calculation of
processes with external SUSY particles. We also briefly discuss the
application of the new {\tt FeynArts} model file to the calculation of
two-loop corrections to cMSSM Higgs boson
masses~\cite{mhiggs2Lp2,mhcMSSM2Latat}, and to the evaluation of
partial decay widths of
SUSY electroweak (EW) particles~\cite{LHCxC,LHCxN,LHCxNprod}.

%%%%%%%%%%%%%%%%%%%%%%%%%%%%%%%%%%%%%%%%%%%%%%%%%%%%%%%%%%%%%%%%%%%%%%%%%%%%%%%
%%%%%%%%%%%%%%%%%%%%%%%%%%%%%%%%%%%%%%%%%%%%%%%%%%%%%%%%%%%%%%%%%%%%%%%%%%%%%%%

\section{Renormalization of the cMSSM}

The tree-level Feynman rules of the MSSM are by now well under
control, where the cMSSM had been included into the 
{\tt FeynArts} package~\cite{FA-MSSM}. 
Concerning the renormalization, however, most
calculations in the past chose a prescription that was tailored to one
specific calculation or even one specific part of the (c)MSSM parameter
space. Since the values of the SUSY parameters realized in nature are
unknown, at the current state
scans over large parts of the cMSSM parameter space are necessary.
Furthermore, many processes have to be evaluated simultaneously. Both
requirements make a {\em complete} renormalization of the cMSSM
necessary that is 
valid over the {\em full} (or at least ``large parts'') of the cMSSM
parameter space. 
Only with such a renormalization at hand fully automated calculations in
the cMSSM will be possible. Evidently, calculations at $n$-loop require
an $n$-loop renormalization, where we will focus on the one-loop case.

The program of the renormalization of all (physical) sectors of the
cMSSM has recently been
completed~\cite{mhcMSSMlong,SbotRen,Stop2decay,Gluinodecay,Stau2decay,LHCxC,LHCxN,LHCxNprod,mssmct} 
%(based on earlier work in the MSSM~\cite{dissHR,hr,dissTF,diplTF}% 
%\footnote{For alternative approaches see \citeres{baro,Baro,onshellCNmasses}.}% 
%
(based on earlier work in the MSSM~\cite{dissHR,hr,dissTF,diplTF}; 
for alternative approaches see \citeres{baro,Baro,onshellCNmasses}% 
) and included as a model file {\tt MSSMCT.mod}~\cite{mssmct} into the
{\tt FeynArts} package.  

In the development of the renormalization 
particular emphasis was put on the requirement that the
one-loop corrections stay ``small'' over the full allowed parameter
range. The renormalization includes the scalar fermion sector,
the remaining colored sector, the chargino/neutralino sector and the
Higgs sector (as well as the SM part of the MSSM). 
In principle this is sufficient to evaluate all currently
relevant processes at the one-loop level. Extensive checks have been
performed to ensure ``stability'' of the higher-order corrections over
large(st) parts of the cMSSM parameter space. These tests include 
scalar top and bottom decays~\cite{SbotRen,Stop2decay}, scalar tau
decays~\cite{Stau2decay}, gluino decays~\cite{Gluinodecay} as well as
non-hadronic chargino~\cite{LHCxC} and neutralino
decays~\cite{LHCxN,LHCxNprod}.  
These evaluations are complete at the one-loop level, including hard and
soft QED and QCD radiation.

\renewcommand\arraystretch{1.25}

%%%%%%%%%%%%%%%%%%%%%% T A B L E %%%%%%%%%%%%%%%%%%%%%%%%%%%%%%%%%%%%%%%%%%%%%
\begin{table}[htb!]
\begin{center}
\begin{tabular}{|l|l|l|l||l|l|l|l|} \hline
leptons & $f = f^\dagger\!$ & field & mass &
sleptons & $f = f^\dagger\!$ & field & mass \\ \hline
$\nu_g$ & & \Code{F[1,\,\Brac{\Vg}]} & \Code{0} &
  $\tilde\nu_g$ & & \Code{S[11,\,\Brac{\Vg}]} & \Code{MSf} \\
$\ell_g$ & & \Code{F[2,\,\Brac{\Vg}]} & \Code{MLE} &
  $\tilde\ell_g^s$ & & \Code{S[12,\,\Brac{\Vs,\Vg}]} & \Code{MSf} \\
\hline
\multicolumn{6}{c}{} \\[-1ex] \hline
\multicolumn{4}{|l||}{quarks} &
\multicolumn{4}{|l|}{squarks} \\ \hline
$u_g$ & & \Code{F[3,\,\Brac{\Vg,\Vo}]} & \Code{MQU} &
  $\tilde u_g^s$ & & \Code{S[13,\,\Brac{\Vs,\Vg,\Vo}]} & \Code{MSf} \\
$d_g$ & & \Code{F[4,\,\Brac{\Vg,\Vo}]} & \Code{MQD} &
  $\tilde d_g^s$ & & \Code{S[14,\,\Brac{\Vs,\Vg,\Vo}]} & \Code{MSf} \\
\hline
\multicolumn{6}{c}{} \\[-1ex] \hline
\multicolumn{4}{|l||}{gauge bosons} &
\multicolumn{4}{|l|}{neutralinos, charginos} \\ \hline
$\gamma$ & yes & \Code{V[1]} & \Code{0} &
  $\tilde\chi_n^0$ & yes & \Code{F[11,\,\Brac{\Vn}]} & \Code{MNeu} \\
$Z$ & yes & \Code{V[2]} & \Code{MZ} &
  $\tilde\chi_c^-$ & & \Code{F[12,\,\Brac{\Vc}]} & \Code{MCha} \\ 
$W^-$ & & \Code{V[3]} & \Code{MW} & & & & \\ \hline
\multicolumn{6}{c}{} \\[-1ex] \hline
\multicolumn{4}{|l||}{Higgs/Goldstone bosons} &
\multicolumn{4}{|l|}{ghosts} \\ \hline
$h$ & yes & \Code{S[1]} & \Code{Mh0} &
	$u_\gamma$ & & \Code{U[1]} & \Code{0} \\
$H$ & yes & \Code{S[2]} & \Code{MHH} &
	$u_Z$ & & \Code{U[2]} & \Code{MZ} \\
$A$ & yes & \Code{S[3]} & \Code{MA0} &
	$u_+$ & & \Code{U[3]} & \Code{MW} \\
$G$ & yes & \Code{S[4]} & \Code{MZ} &
	$u_-$ & & \Code{U[4]} & \Code{MW} \\
$H^-$ & & \Code{S[5]} & \Code{MHp} &
	$u_g$ & & \Code{U[5,\,\Brac{\Vu}]} & \Code{0} \\
$G^-$ & & \Code{S[6]} & \Code{MW} & & & & \\ \hline
\multicolumn{6}{c}{} \\[-1ex] \hline
\multicolumn{4}{|l||}{gluon} &
\multicolumn{4}{|l|}{gluino} \\ \hline
$g$ & yes & \Code{V[5,\,\Brac{\Vu}]} & \Code{0} &
$\tilde g$ & yes & \Code{F[15,\,\Brac{\Vu}]} & \Code{MGl} \\
\hline
\end{tabular}
\caption{\label{tab:mssmparticles}The particle content of
\Code{MSSMCT.mod}.}
\end{center}
\end{table}
%%%%%%%%%%%%%%%%%%%%%% T A B L E %%%%%%%%%%%%%%%%%%%%%%%%%%%%%%%%%%%%%%%%%%%%%

%%%%%%%%%%%%%%%%%%%%%% T A B L E %%%%%%%%%%%%%%%%%%%%%%%%%%%%%%%%%%%%%%%%%%%%%
\begin{table}[htb!]
%\vspace{-1em}
\begin{minipage}[t]{.5\hsize}
\begin{alignat*}{2}
\Vg &= \Code{Index[Generation]} &&= 1\dots 3\,, \\
\Vo &= \Code{Index[Colour]}     &&= 1\dots 3\,, \\
\Vu &= \Code{Index[Gluon]}      &&= 1\dots 8\,, \\
\Vs &= \Code{Index[Sfermion]}   &&= 1\dots 2\,, \\
\Vn &= \Code{Index[Neutralino]} &&= 1\dots 4\,, \\
\Vc &= \Code{Index[Chargino]}   &&= 1\dots 2\,. \\
\end{alignat*}
\end{minipage}\begin{minipage}[t]{.5\hsize}
\begin{gather*}
\text{(S)fermions are indexed by\qquad\qquad} \\
t = \begin{cases}
  1 & \text{(s)neutrinos}, \\
  2 & \text{charged (s)leptons}, \\
  3 & \text{up-type (s)quarks}, \\
  4 & \text{down-type (s)quarks}.
\end{cases}
\end{gather*}
\end{minipage}
\vspace{-1em}
\caption{\label{tab:indices}Index labels and ranges used in 
{\tt MSSMCT.mod}.}
%\vspace{1em}
\end{table}
%%%%%%%%%%%%%%%%%%%%%% T A B L E %%%%%%%%%%%%%%%%%%%%%%%%%%%%%%%%%%%%%%%%%%%%%
\renewcommand\arraystretch{1.2}

%%%%%%%%%%%%%%%%%%%%%% T A B L E %%%%%%%%%%%%%%%%%%%%%%%%%%%%%%%%%%%%%%%%%%%%%
\renewcommand\arraystretch{1.13}
\begin{table}[htb!]
\begin{center}
\begin{tabular}{|l|l|}
\hline
\Code{Mh0}, \Code{MHH}, \Code{MA0}, \Code{MHp} &
	Higgs masses $\Mh$, $\MH$, $\MA$, $\MHp$ \\
\Code{Mh0tree}, \Code{MHHtree}, \Code{MA0tree}, \Code{MHptree} &
	tree-level Higgs masses
%	$m_h^{(0)}$, $m_H^{(0)}$, $\MA^{(0)}$, $\MHp^{(0)}$
	\\
\Code{TB}, \Code{CB}, \Code{SB}, \Code{C2B}, \Code{S2B} &
	$\tan\beta$, $\cos\beta$, $\sin\beta$,
	$\cos 2\beta$, $\sin 2\beta$ \\
\Code{CA}, \Code{SA}, \Code{C2A}, \Code{S2A} &
	$\cos\alpha$, $\sin\alpha$,
	$\cos 2\alpha$, $\sin 2\alpha$
	(tree-level $\alpha$) \\
\Code{CAB}, \Code{SAB}, \Code{CBA}, \Code{SBA} &
	$\cos(\alpha + \beta)$, $\sin(\alpha + \beta)$,
	$\cos(\beta - \alpha)$, $\sin(\beta - \alpha)$ \\
\Code{MUE} &
	Higgs-doublet mixing parameter $\mu$ \\ 
\hline
\Code{MGl} &
	gluino mass $\mgl$ \\
\Code{SqrtEGl} &
	root of the gluino phase, $\re^{\ri\phigl/2}$ \\
\Code{MNeu[\Vn]} &
	neutralino masses $m_{\neu{n}}$ \\
\Code{ZNeu[\Vn,\,\Vnp]} &
	neutralino mixing matrix $\matr{N}_{nn'}$ \\
\Code{MCha[\Vc]} &
	chargino masses $\mcha{c}$ \\
\Code{UCha[\Vc,\,\Vcp]}, \Code{VCha[\Vc,\,\Vcp]} &
	chargino mixing matrices $\matr{U}_{cc'}, \matr{V}_{cc'}$ \\ 
\hline
\Code{MSf[\Vs,\,\Vt,\,\Vg]} &
	sfermion masses $m_{\Sf_{t,sg}}$ \\
\Code{USf[\Vt,\,\Vg][\Vs,\,\Vsp]} &
	sfermion mixing matrix $U^{\Sf_{tg}}_{ss'}$ \\
\Code{Af[\Vt,\,\Vg,\,\Vgp] } &
	soft-breaking trilinear $A$-parameters
	$\mati{\matr{A}_{f_t}}_{gg'}$ \\ 
\hline
\Code{MW}, \Code{MZ} &
	gauge-boson masses $\MW$, $\MZ$ \\
\Code{Mf[\Vt,\Vg]} &
        fermion masses $\mf{tg}$ \\
\hline
\Code{CW}, \Code{SW} &
	$\CW\equiv\cos\theta_{\text{w}} = \MW/\MZ$,
	$\SW\equiv\sin\theta_{\text{w}}$ \\
\Code{EL} &
	electromagnetic coupling constant $e$ \\
\Code{GS} &
	strong coupling constant $g_s$ \\
\hline
\end{tabular}
\caption{\label{tab:modelsyms}Symbols representing the SM and MSSM parameters 
in \Code{MSSMCT.mod}. }
\end{center}
\end{table}
%%%%%%%%%%%%%%%%%%%%%% T A B L E %%%%%%%%%%%%%%%%%%%%%%%%%%%%%%%%%%%%%%%%%%%%%

\medskip
To give an idea about the new model file we briefly review the various
masses, coupling constants etc.\ as well as the respective counterterms
implemented into \Code{MSSMCT.mod} (all details can be found in
\citere{mssmct}). In \refta{tab:mssmparticles} we list the particle
content of the model file, where the respective index ranges are given
in \refta{tab:indices}. The symbols for the masses of the particles, for
couplings and mixing angles are
shown in \refta{tab:modelsyms}. Within the Higgs boson sector the
tree-level masses are taken distinct from the higher-order corrected
masses (which can be obtained via the automatic link to 
{\tt FeynHiggs}~\cite{feynhiggs,mhiggslong,mhiggsAEC,mhcMSSMlong,Mh-logresum}). 
$\tb$ denotes the ratio of the two vacuum expectation values, and $\al$
is the angle that diagonalizes the (tree-level) $\cp$-even Higgs
sector. 
Within the cMSSM all three neutral Higgs bosons can mix to give rise to
three $\cp$-mixed states, $h_i$ ($i = 1,2,3$). 
When composing a vertex $\Gamma_{h_i}$ from the 
corresponding tree-level amplitudes $\Gamma_h$, $\Gamma_H$, and 
$\Gamma_A$, a set of finite $Z$-factors is needed to ensure
correct on-shell properties of the external Higgs boson
$h_i$~\cite{mhcMSSMlong}, 
\begin{equation}
\label{eq:zfactors123}
\Gamma_{h_i} = \hat Z_{i1}\Gamma_h + \hat Z_{i2}\Gamma_H +
  \hat Z_{i3}\Gamma_A + \ldots\,,
\end{equation}
where the ellipsis represents contributions from the mixing with the 
Goldstone and $Z$ boson. %, which have to be taken into account explicitly.  
The $Z$-factor matrix $\hat Z_{ij}\equiv\Code{ZHiggs[\Vi,\,\Vj]}$ is not 
in general unitary.  Its lower $3\times 3$ part is computed by {\tt FeynHiggs}
and application at the amplitude level automatically takes any 
absorptive contribution into account.  Technically this is most easily 
accomplished using the FeynArts add-on model file
\Code{HMix.mod}~\cite{SUSY07} which mixes 
$h = \Code{S[1]}$, $H = \Code{S[2]}$, and 
$A = \Code{S[3]}$ into two variants of the loop-corrected states $h_i$,
\begin{subequations}
\label{eq:hmix}
\begin{alignat}{2}
\Code{S[0,\,\Brac{\Vi}]} &=
	\sum_{j = 1}^3 \Code{UHiggs[\Vi,\,\Vj]}~\Code{S[\Vj]}\,, &
	&\qquad\begin{minipage}[t]{.4\hsize}
	\raggedright
	with unitary \Code{UHiggs} (no absorptive \\
	contrib.), for use on internal lines,
	\end{minipage} \\
\Code{S[10,\,\Brac{\Vi}]} &=
	\sum_{j = 1}^3 \Code{ZHiggs[\Vi,\,\Vj]}~\Code{S[\Vj]}\,, &
	&\qquad\text{inserted only on external lines.}
\end{alignat}
\end{subequations}

%%%%%%%%%%%%%%%%%%%%%% T A B L E %%%%%%%%%%%%%%%%%%%%%%%%%%%%%%%%%%%%%%%%%%%%%
\begin{table}[htb!]
\begin{center}
\begin{tabular}{|l|l|}
\multicolumn{2}{l}{Higgs-boson Sector} \\
\hline
\Code{dZ\bbarb Higgs1[\Vh,\,\Vhp]} &
	Higgs field RCs \\
\Code{dMHiggs1[\Vh,\,\Vhp]} &
	Higgs mass RCs \\
\Code{dTh01}, \Code{dTHH1}, \Code{dTA01} &
	Higgs tadpole RCs \\
\Code{dZH1}, \Code{dZH2},
\Code{dTB1}, \Code{dSB1}, \Code{dCB1} &
	RCs related to $\beta$ \\
\hline
\multicolumn{2}{l}{Gauge-boson Sector } \\
\hline
\Code{dMZsq1}, \Code{dMWsq1} &
	gauge-boson mass RCs \\
\Code{dZAA1}, \Code{dZAZ1}, \Code{dZZA1}, \Code{dZZZ1}, \Code{dZ\bbarb W1} &
	gauge-boson field RCs \\
\Code{dSW1}, \Code{dZe1} &
	coupling-constant RCs \\
\hline
\multicolumn{2}{l}{Chargino/Neutralino Sector } \\
\hline
\Code{dMCha1[\Vc,\,\Vcp]} &
	chargino mass RCs \\
\Code{dMNeu1[\Vn,\,\Vnp]} &
	neutralino mass RCs \\
\Code{dMino11}, \Code{dMino21}, \Code{dMUE1} &
	RCs for $M_1$, $M_2$, $\mu$ \\
\Code{dZ\bbarb f\opt{L}{R}1[12,\,\Vc,\,\Vcp]} &
	chargino field RCs \\
\Code{dZ\bbarb f\opt{L}{R}1[11,\,\Vn,\,\Vnp]} &
	neutralino field RCs \\
\hline
\multicolumn{2}{l}{Fermion Sector } \\
\hline
\Code{dMf1[\Vt,\,\Vg]} &
	fermion mass RCs \\
\Code{dZ\bbarb f\smash{\opt{L}{R}}1[\Vt,\,\Vg,\,\Vgp]} &
	fermion field RCs \\
\Code{dCKM1[\Vg,\,\Vgp]} &
	CKM-matrix RCs \\
\hline
\multicolumn{2}{l}{Squark Sector } \\
\hline
\Code{dMSfsq1[\Vs,\,\Vsp,\,3\rI4,\,\Vg]} &
	squark mass RCs \\
\Code{dAf1[3\rI 4,\,\Vg,\,\Vg]} &
	trilinear squark coupling RCs \\
\Code{dZ\bbarb Sf\opt{L}{R}1[\Vs,\,\Vsp,\,3\rI4,\,\Vg]} &
	squark field RCs \\
\hline
\multicolumn{2}{l}{Slepton Sector } \\
\hline
\Code{dMSfsq1[\Vs,\,\Vsp,\,1\rI 2,\,\Vg]} &
	slepton mass RCs \\
\Code{dAf1[2,\,\Vg,\,\Vg]} &
	trilinear slepton coupling RCs \\
\Code{dZ\bbarb Sf\opt{L}{R}1[\Vs,\,\Vsp,\,1\rI 2,\,\Vg]} &
	slepton field RCs \\
\hline
\multicolumn{2}{l}{Gluino Sector } \\
\hline
\Code{dMGl1} &
	gluino mass RC \\
\Code{dZ\bbarb Gl\opt{L}{R}1} &
	gluino field RCs \\
\hline
\multicolumn{2}{l}{Gluon Sector } \\
\hline
\Code{dZgs1} &
	strong-coupling-constant RC \\
\Code{dZGG1} &
	gluon field RCs \\
\hline
\end{tabular}
\caption{\label{tab:ct}RCs used in 
\Code{MSSMCT.mod}, where $a$\rI$b$ means `$a$ or $b$' and
\Code{dZ\bbarb} stands for both \Code{dZ} and \Code{dZbar} (see text).}
\end{center}
\end{table}
%%%%%%%%%%%%%%%%%%%%%% T A B L E %%%%%%%%%%%%%%%%%%%%%%%%%%%%%%%%%%%%%%%%%%%%%

\noindent
More details can be found in \citeres{mhcMSSMlong,mssmct}.
The renormalization constants (RCs) are listed in \refta{tab:ct}. For the
Higgs boson sector, besides the mass and field renormalization constants
also the tadpole counterterms are listed, which correspond to the terms
linear in the Higgs fields in the Higgs potential. Several field
renormalization constants are given in a barred and unbarred version,
differentiating between incoming/outgoing (anti)particles, see
\citeres{Stop2decay,LHCxC,LHCxN,mssmct} for more details.
Several electroweak SM RCs have been omitted (such as \Code{dMW1} etc.),
since they are defined identical to the pure SM case. 
The {\tt FeynArts}/{\tt FormCalc} framework provides a default
implementation of the determination of all RCs, where again the details
can be found in
\citeres{mhcMSSMlong,SbotRen,Stop2decay,Gluinodecay,Stau2decay,LHCxC,LHCxN,LHCxNprod,mssmct}. 
It should be stressed again that this default implementation is based on
the requirement of ``stability'' of the higher-order corrections over
large(st) parts of the cMSSM parameter space.

%%%%%%%%%%%%%%%%%%%%%%%%%%%%%%%%%%%%%%%%%%%%%%%%%%%%%%%%%%%%%%%%%%%%%%%%%%%%%%%
%%%%%%%%%%%%%%%%%%%%%%%%%%%%%%%%%%%%%%%%%%%%%%%%%%%%%%%%%%%%%%%%%%%%%%%%%%%%%%%

\section{Example applications}

In this section we briefly review some of the yet existing applications of 
\Code{MSSMCT.mod}.

%%%%%%%%%%%%%%%%%%%%%%%%%%%%%%%%%%%%%%%%%%%%%%%%%%%%%%%%%%%%%%%%%%%%%%%%%%%%%%%

\subsection{Two-loop corrections to Higgs boson masses}

\begin{figure}[htb!]
\centering
\includegraphics[width=0.62\textwidth]{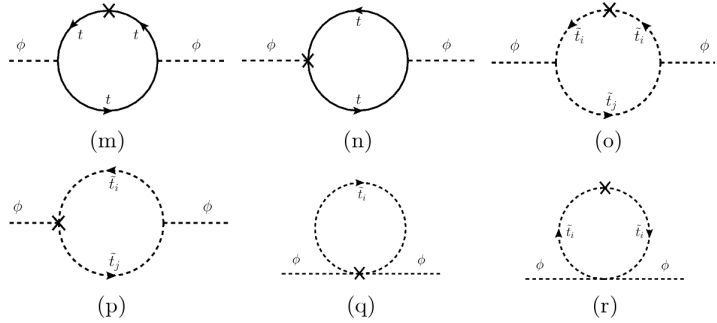}
\caption{
Generic one-loop contributions to cMSSM Higgs-boson self-energies with
sub-loop renormalization; $\phi = h, H, A$; $t$ denotes the top quark;
$\Stop_{i,j}$ the scalar tops with $i,j = 1,2$.
}
\label{fig:asat-selfenergies}
\end{figure}

Two-loop corrections to cMSSM Higgs boson masses, obtained in the
Feynman-diagrammatic approach, require the calculation
of two-loop Higgs-boson self-energies, which in turn require a
renormalization of the Higgs boson sector at the two-loop level (see
\citeres{mhiggslong,mhiggs2Lasab,mhcMSSMlong,mhcMSSM2L,mhiggs2Lp2,mhcMSSM2Latat,mhiggsEP2L}). 
Furthermore, a sub-loop renormalization of the corresponding one-loop
diagrams is necessary. Consequently, a full two-loop calculation of the
Higgs-boson self-energies requires a full renormalization of the cMSSM
at the one-loop level. Recently, two new two-loop calculations were
presented. One consists of the \order{\alt^2} contributions involving
complex phases~\cite{mhcMSSM2Latat}, the other one of the momentum
dependent two-loop part of the \order{\alt\als} corrections for real
parameters~\cite{mhiggs2Lp2}. Both types of corrections can yield
contributions to $\Mh$ larger than the current experimental
uncertainty. Corresponding one-loop diagrams with sub-loop
renormalization are depicted in \reffi{fig:asat-selfenergies} (taken
from \citere{mhiggs2Lp2}).

%%%%%%%%%%%%%%%%%% F I G U R E %%%%%%%%%%%%%%%%%%%%%%%%%%%%%%%%%%%%%%%%%%%%%%%%
%%%%%%%%%%%%%%%%%% F I G U R E %%%%%%%%%%%%%%%%%%%%%%%%%%%%%%%%%%%%%%%%%%%%%%%%

%%%%%%%%%%%%%%%%%%%%%%%%%%%%%%%%%%%%%%%%%%%%%%%%%%%%%%%%%%%%%%%%%%%%%%%%%%%%%%%

\subsection{Decays of electroweak SUSY particles}

\newcommand{\Sg}{\ensuremath{{S}_g}}
\newcommand{\Sh}{\ensuremath{{S}_h}}
\newcommand{\SgII}{\ensuremath{{\tilde S}_g}}
\newcommand{\ShII}{\ensuremath{{\tilde S}_h}}
\newcommand{\Al}{A_l}

The second example concerns one-loop processes with external SUSY
particles. A precise prediction of, e.g., SUSY production cross sections
and decay branching ratios is necessary to obtain reliable bounds on the
MSSM parameter space from LHC SUSY searches, to correctly interpret any
possible signal at the LHC and to exploit the potential of a future
$e^+e^-$ collider such as the ILC, where measurements at the per-cent
level will be possible.  

Calculations of decay widths
of SUSY particles in the cMSSM, using \Code{MSSMCT.mod} (and its
earlier versions) have been published in 
\citeres{SbotRen,Stop2decay,Gluinodecay,Stau2decay,LHCxC,LHCxN,LHCxNprod}. 
Here we take one representative example from \citere{LHCxN} that
involves both a Higgs particle and a Dark Matter particle in the final
state, the decay of the fourth neutralino to the lightest neutralino and
the lightest Higgs boson, $\neu4 \to \neu1 \He$. The parameters are
given in \refta{tab:para}. 
$\MTwo$ (the $SU(2)$ soft SUSY-breaking parameter) and $\mu$ (the Higgs
mixing parameter) are chosen such that the values for $\mcha{1}$ and
$\mcha{2}$ are fulfilled. Here the ambiguity in the hierarchy of $\MTwo$
and $\mu$ results in two scenarios: $\mu > \MTwo$ yields a
higgsino-like $\neu4$, denoted as \Sh; $\mu < \MTwo$ gives a gaugino-like
$\neu4$, denoted as \Sg.
$|\MOne|$ (the absolute value of the $U(1)$ soft SUSY-breaking
parameter) is obtained from  
$|\MOne| = \frac{5}{3} \tan^2 \thw \MTwo \approx \edz \MTwo$, 
where $\phiMe$ is kept as a free parameter.

%%%%%%%%%%%%%%%%%%%%% T A B L E %%%%%%%%%%%%%%%%%%%%%%%%%%%%%%%%%%%%%%%%%%%%%%
\begin{table}[ht!]
\renewcommand{\arraystretch}{1.2}
\BC
\begin{tabular}{|c|c|c|c|c|c|c|c|c|c|c|}
\hline
$\tb$ & $\MHp$ & $\mcha{2}$ & $\mcha{1}$ 
& $\MslL$ & $\MslR$ & $\Al$ 
& $\MsqL$ & $\MsqR$ & $\Aq$
\\ \hline\hline
$ 20$ & $ 160$ & $ 600$ & $ 350$ & $ 300$ & $ 310$ & $ 400$ 
& $ 1300$ & $ 1100$ & $ 2000$
\\ \hline
\end{tabular}
\caption{MSSM parameters with all 
  mass parameters are in$\gev$. $\MHp$ denotes the mass of the charged
  Higgs boson, $\mcha{1,2}$ are the chargino masses, $\MslL$ and $\MslR$
  are the diagonal entries in the slepton mass matrices (taken to be
  universal for the three generations), $\Al$ is the trilinear
  Higgs-slepton coupling; $\MsqL$, $\MsqR$ and $\Aq$ are the
  corresponding squark sector parameters (see \citere{LHCxN} for details).
}
\label{tab:para}
\EC
\renewcommand{\arraystretch}{1.0}
\vspace{-1em}
\end{table}
%%%%%%%%%%%%%%%%%%%%% T A B L E %%%%%%%%%%%%%%%%%%%%%%%%%%%%%%%%%%%%%%%%%%%%%%

\begin{figure}[htb!]
\centering
\includegraphics[width=0.45\textwidth,height=5cm]{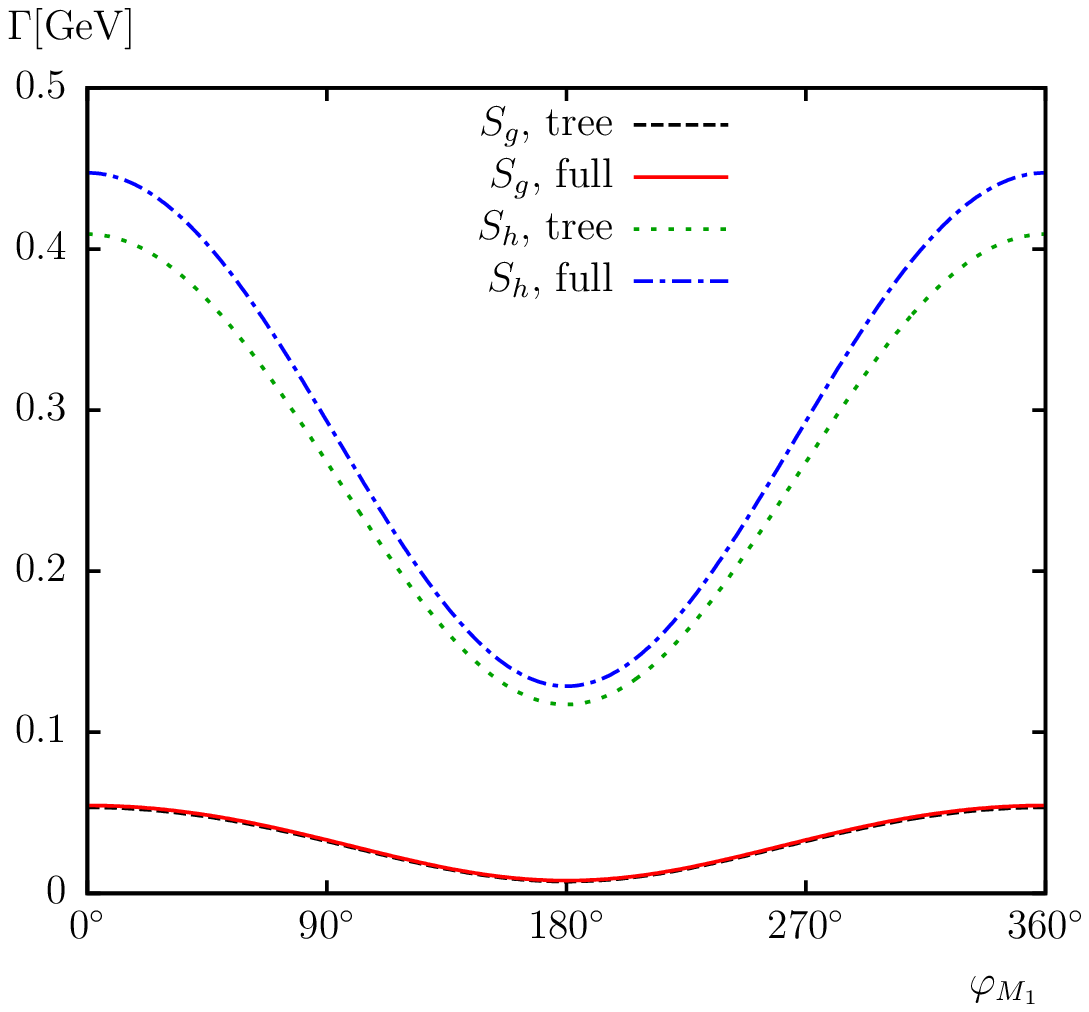} \quad
\includegraphics[width=0.45\textwidth,height=5cm]{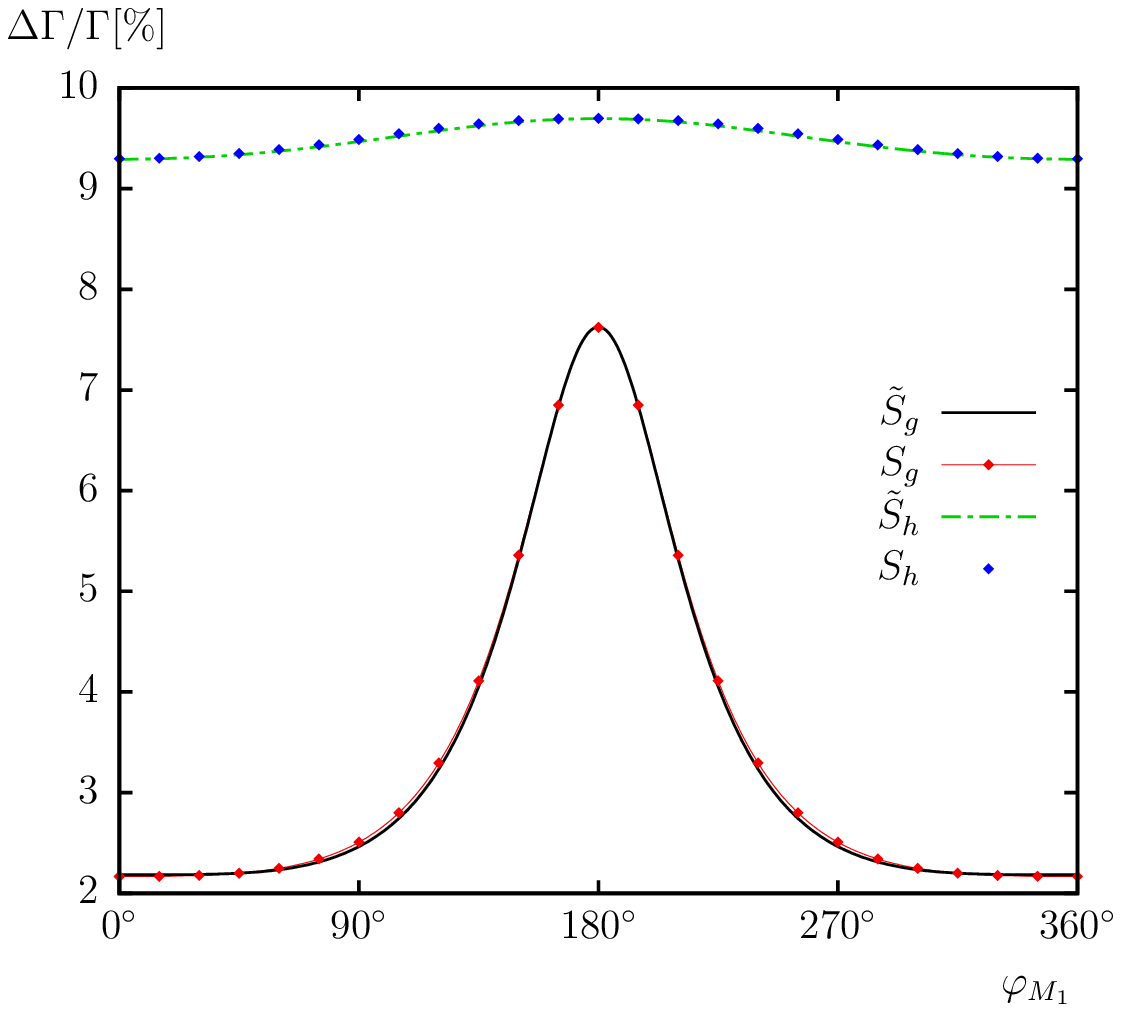}
\caption{Left: absolute values of $\Ga(\neu4 \to \neu1 \He)$ in the
  scenarios \Sg\ and \Sh\ (see text) at the tree-level and the full
  one-loop level as a function of $\phiMe$.
  Right: relative size of the one-loop corrections in our default
  renormalization scheme (\Sg\ and \Sh) and in an alternative scheme
  (\SgII\ and \ShII, see text) as a function of $\phiMe$.
}
\label{fig:neu4neu1h1}
\end{figure}

The absolute size of $\Ga(\neu4 \to \neu1 \He)$ in \Sg\ and \Sh\ is
shown in the left plot of \reffi{fig:neu4neu1h1} as a function of
$\phiMe$; separately shown are the tree-level results and the full
one-loop calculation. A strong dependence of the absolute value of the
decay width in both scenarios at the tree-level and at the one-loop
level on this phase can be observed. The right plot shows the relative
size of the one-loop 
corrections. Again a strong dependence of the size of those corrections
on the phase of $\MOne$ can be seen. Besides the default implementation
of our renormalization scheme, denoted as \Sg\ and \Sh, we also show the
results in an alternative on-shell scheme~\cite{LHCxN,ChaNeuRenII} that
differs in the treatment of the complex phases, denoted as \SgII\ and
\ShII. It can be observed that the two schemes agree for real parameters
($\phiMe = 0, \pi$) and show small differences (indicating the size of
respective two-loop corrections) for complex parameters at or below the
per-cent level.

%%%%%%%%%%%%%%%%%% F I G U R E %%%%%%%%%%%%%%%%%%%%%%%%%%%%%%%%%%%%%%%%%%%%%%%%
%%%%%%%%%%%%%%%%%% F I G U R E %%%%%%%%%%%%%%%%%%%%%%%%%%%%%%%%%%%%%%%%%%%%%%%%

%%%%%%%%%%%%%%%%%%%%%%%%%%%%%%%%%%%%%%%%%%%%%%%%%%%%%%%%%%%%%%%%%%%%%%%%%%%%%%%
%%%%%%%%%%%%%%%%%%%%%%%%%%%%%%%%%%%%%%%%%%%%%%%%%%%%%%%%%%%%%%%%%%%%%%%%%%%%%%%

\subsubsection*{Acknowledgements}

S.H.\ thanks the organizers of L\&L\,2014 for the invitation and the (as
always!) inspiring atmosphere, and for being flexible with the talk
asignments. 
We thank
A.~Bharucha
with whom some of the results presented here have been obtained.
The work of S.H.\ is supported 
in part by the 
Spanish MICINN's Consolider-Ingenio 2010 Program under grant MultiDark
CSD2009-00064. 
We furthermore thank the GRID computing network at IFCA for 
technical help with the OpenStack cloud
infrastructure~\cite{openstack}, where many of the numerical results
shown here have been obtained.

%%%%%%%%%%%%%%%%%%%%%%%%%%%%%%%%%%%%%%%%%%%%%%%%%%%%%%%%%%%%%%%%%%%%%%%%%%%%%%%
%%%%%%%%%%%%%%%%%%%%%%%%%%%%%%%%%%%%%%%%%%%%%%%%%%%%%%%%%%%%%%%%%%%%%%%%%%%%%%%

\end{document}